# Model Risk Management for Generative AI In Financial Institutions

**Authors**

Anwesha Bhattacharyya, Ye Yu, Hanyu Yang, Rahul Singh, Tarun Joshi, Jie Chen, Kiran Yalavarthy

Wells Fargo

Abstract

The success of OpenAI's ChatGPT in 2023 has spurred financial enterprises into exploring Generative AI applications to reduce costs and/or drive revenue within different lines of businesses in the Financial Industry. While these applications offer strong potential for efficiencies, they introduce new model risks, primarily hallucinations and toxicity. As highly regulated entities, financial enterprises (primarily large US banks) are obligated to enhance their model risk framework with additional testing and controls to ensure safe deployment of such applications. This paper outlines the key aspects for model risk management of generative AI model with a special emphasis on additional practices required in model validation.

## 1 Introduction

Large Language Models (LLM) and Generative Artificial Intelligence (GenAI) models represent a significant advancement in artificial intelligence, characterized by their ability to generate new data that resembles existing data. These models, which include architectures such as Generative Adversarial Networks (GANs), Variational Autoencoders (VAEs), and Transformers have gained attention for their ability to perform complex tasks across various domains. Pre-trained LLMs (Foundational Models: FM) can potentially understand large volumes of unstructured data and can be adopted to generate insightful content for a variety of business use cases. In financial institutions, these models are being leveraged to streamline operations, improve decision-making processes, and enhance overall efficiency. Applications range from trading and risk assessment to customer engagement and personalized financial services. By harnessing the power of GenAI, financial organizations can not only automate routine tasks but also gain deeper insights into market dynamics and customer behaviors.

In risk management, GenAI models can play a pivotal role in scanning transactions with other institutions, market news, asset prices and other proprietary information and assets to provide customized solutions to risk management (Agarwal, et al. 2024). This capability helps streamline compliance processes, ensuring that organizations adhere to regulatory requirements while optimizing capital allocation. Additionally, GenAI is being utilized to enhance customer service through advanced chatbots and virtual assistants. By analyzing customer inquiries and behavioral data, these AI-driven systems can provide tailored responses, improving customer experience and satisfaction, thereby increasing customer retention rates. While the benefits of GenAI in enhancing operational efficiency are evident, organizations must also consider a) cost in terms of computational power and human talent, b) heightened risk from the ethical implications and potential biases inherent in AI systems and c) potential vulnerabilities to malicious exploitation. As such, the regulatory landscape is rapidly evolving, suggesting restrictions and/or prescriptions towards safe use of GenAI models,[ (The European Parliament and The Council of The European Union 2024), (National Institute of Standards and Technology 2024)]. Therefore, an enhanced and an

*****This material represents the views of the authors and does not necessarily reflect those of Wells Fargo.*

evolving model risk management framework with additional controls is required to support the successful implementation of GenAI in financial institutions.

Typically, in financial institutions the model risk management is a holistic approach where the responsibilities are shared across by the line of business, model development team, and the model risk management team [(Board of Governors of the Federal Reserve System and Office of the Comptroller of the Currency 2011), also referred as SR11-7]. In this paper, we primarily focus on the responsibilities of the model risk management team.

The unique contribution of this paper is providing an end-to-end model risk management framework for GenAI models aligned to existing SR11-7 regulations. To best of our knowledge, this is the first paper that defines the incremental testing required for effective model risk management of GenAI models along the three pillars of SR-11-7: conceptual soundness, outcome analysis, and ongoing monitoring. Additionally, when appropriate the paper suggests methodologies that are being or may be leveraged to implement this testing.

The rest of the paper is structured as follows: Section 2 discusses common applications of LLM/GenAI within financial institutions; their heightened risk and the different stages of model lifecycle process that are leveraged to quantify, mitigate and monitor these risks; Section 3 proposes an end-to-end GenAI model risk framework including all salient pillars for SR-117 with examples when necessary. We conclude in Section 4.

## 2 GenAI Usage and Heightened Risk

### Applications in Financial Institutions

Based on the business use, LLMs may be modeled to generate a bounded output (example: classification, information retrieval, or extractive answers) or may be modeled to generate new text (example: summarization or retrieval augmented generation) conditional to their understanding of the input text. In the context of this paper, we primarily focus on the generative use cases with a few examples listed below.

Table 2-1 GenAI usage categories and examples

| GenAI usage | Description of usage | Application examples in banks |
|---|---|---|
| Generative Summarization | Summarizing information from a specific provided context. | Model designed to enhance efficiency and effectiveness of the complaint resolution process. For example, summarize customer phone call transcripts to elicit insights around customer pain points. |
| Retrieval Augmented Generation (RAG) | Generating summarized information based on relevant retrieved information for an input query. | <ul><li>Chatbots for retrieving and summarizing third party research for the purpose of creating credit memos.</li><li>Tools to retrieve and summarize internal policy documents.</li></ul> |
| General Content Generation | General purpose content generation using open ended prompts (queries) to LLMs. | Generating first drafts for internal communication, marketing, and internal legal or policy documents. |

## Heightened Risks for Generative AI Models

Use of GenAI models is associated with heightened or novel risks compared to general quantitative or Machine Learning based models. These risks can be classified as model risk and non-model risk and are described in Table 2-2.

**Table 2-2 Heightened Risk for GenAI Applications**

| | GenAI Heightened Risk | Description |
|---|---|---|
| **GenAI Model Risk** | Data and Privacy risk | • It is virtually impossible to test the data integrity for FM as they are trained on petabytes of data. Moreover, customizing an FM for a domain specific use runs the risk of exposing propriety and confidential data. (Duffourc, Gerke and Kollnig 2024) |
| | Explainability | • These are highly complex models with billions of parameters. The black-box nature with complex architecture, along with lack of structured input and output makes it challenging to explain model outcomes and decisions. |
| | Performance and Hallucination | • A GenAI model may produce content that is factually incorrect leading to a misinterpretation of facts from a model user's perspective. Worse, the factual inaccuracies may be subtle, where the generation captures the trend but grossly misinterprets the details especially where financial numbers are involved. This is typically known as hallucination risk.<br>• Performance evaluation for input text and output text pair is not straightforward and requires either carefully designed metrics or expensive human annotations.<br>• The scope of freeform text query allows for large variation of model instructions and subsequent variation in model output and performance for the same intended task. |
| | Fairness/Toxicity | • A GenAI model may produce content that is factually correct but might be laden with inappropriate language. Worse, the generated content may be laced with deep rooted racial or gender biases emanating from its pre-trained datasets. |
| | Usage risk | • An out-of-box FM can be used for more than one purpose, leading to heightened usage risk beyond approved use cases. |
| **Non-Model Gen AI Risks** | Reputation Risk | • Use case may lead to heightened reputation risk if it results in a change to customer servicing needs, triggers a negative customer experience or alters the terms of a customer's existing relationship with the institution. |
| | Regulatory, Legal, and Compliance Risk | • The use case may introduce innovations that are new to the industry or directly impact consumers. This may lead to additional regulatory, legal and compliance risk. |
| | Third-Party Risks | • Often GenAI models are vendor models or in house models hosted by third-party platforms and thus may invite third-party risks.<br>• The model may utilize third-party copyrighted or otherwise legally protected documents.<br>• Similarly, the model may enhance risk of data leakage if a third party such as cloud service provider is involved in data processing or model components require technological support from external third-party providers. |
| | Technology Risk | • GenAI models require sophisticated technological support that may induce change in existent critical or high rated regularly used applications |
| | Cyber-security Risk | • Vulnerabilities in the GenAI architecture or access of the platform through internet or mobile devices may lead to breaches and be exploited for malicious attacks and fraudulent purposes. |
| | Human Capital Risk | • GenAI applications targeting operational efficiency may impact team structure and cause a decrease in human capital requirement whereas in some cases implementation and ongoing support may result in increased human capital requirement. |

The mitigation of the aforementioned heightened risks requires additional guardrails on the implementation of these models and addressing these issues through various stages of the model risk management framework. Failure in successful mitigation of these risks may result in legal, regulatory, reputational and/or financial consequences.

## Model Lifecycle

To understand the various stages where model risk management plays a role in mitigating the heightened risks associated with GenAI models, it is important to get an overview of the model lifecycle which is represented in Figure 2-1.

- The lifecycle is initialized with identification of a use case. This is followed by a thorough risk assessment of all inherent risks of model use including all heightened risk by multiple independent risk assessment bodies.
- Model developers act as the first line of defense, testing the GenAI model comprehensively on different aspects. This may entail choice of foundational model, developing an evaluation framework along multiple dimensions of completeness, relevance, correctness, and alignment to test for heightened GenAI risks. Developers also test adequacy of implementation plan that optimizes the modeling choices (such as prompting vs fine-tuning) and post-model compression techniques (such as pruning, quantization and distillation) to meet the available computing budget.
- Validation acts as a second line of defense that reviews the challenges and results from developers and conducts independent testing for conceptual soundness and outcome analysis. The second line also typically tests implementation consistency in the production vs development environment, presence of mandatory controls to mitigate heightened risks and concreteness of monitoring plan based on which implementation approval is given.
- At model deployment stage, procedural checks are run to ensure model outputs are as expected and the necessary controls for mitigating risks related to model use are locked down. As the model is used, the presence and effectiveness of these controls are periodically tested.
- Finally, model performance is monitored periodically to evaluate performance deterioration in production and any increase in model use related risks. The second line periodically reviews the monitoring as well.

**Figure 2-1 Model Lifecycle Procedure**

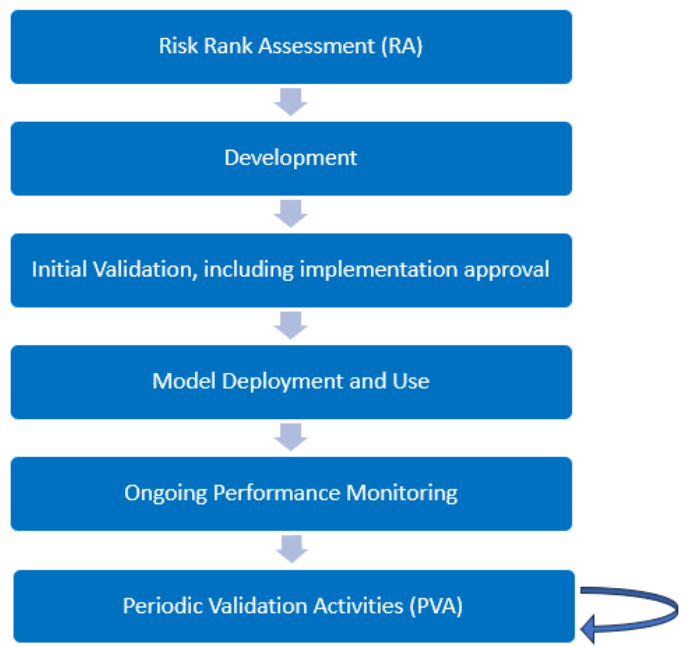

# 3 Model Risk Management for GenAI models.

## Initial Risk Assessment

The inherent risk of a model is directly tied to the model use and the general principles of risk assessment are also applicable to any GenAI use case. Based on model usage, a model is assigned a risk rating that reflects the risk to the financial institution and determines the minimum standards of model risk management activities required during the model lifecycle.

The risk rating of a model is supported by a comprehensive understanding of the model use, model users and the business processes supported by the model. The final risk rating for a model should be determined by the reliance of the business on the model output, the material impact on the business from model errors, the complexity of the modeling choices, and the feasibility of the implementation of required controls for operating the model. It should be noted that the material impact on business might require assessment along different dimensions including financial impact, customer/reputational impact, and regulatory impact emanating from model use and model user.

## Model Development and Validation Standards.

Per SR-117 (Board of Governors of the Federal Reserve System and Office of the Comptroller of the Currency 2011), model risk management is shared between model developers and model validation. This section provides an outline of the required testing from model validation's perspective. We note that the ideas presented here can be applied at model development phase as well.

SR11-7 prescribes model validation as the set of processes and activities intended to verify that models are performing as expected, in line with their design objectives and business uses. Effective model validation should help reduce model risk by identifying appropriate use, model errors, corrective actions, and provide information about the source and extent of model risk accordingly.

Model validation focuses on the key components of conceptual soundness (CS) and outcome analysis (OA). Validation may also include testing of model implementation and review of performance monitoring plans which are discussed in the subsequent sections. For each key testing component, we propose several focus areas to be considered and tested, together with the rationale of the proposed tests. Table 31 summarizes these requirements.

Table 3-1 LLM/GenAI model testing framework and the testing rationale.

| Testing components | LLM/GenAI-specific challenges/weakness/vulnerabilities | Focus areas |
|---|---|---|
| Conceptual Soundness (CS) | **CS #1** - LLM/GenAI is a quickly evolving domain and is usually open-source/vendor-based with limited developmental details shared. | **CS Review #1** - LLM literature review and selection rationale |
| | **CS #2** – Models are based on unstructured data which often require manual labeling for performance assessment. Leakage of sensitive information contained in the development and production data is a concern. | **CS Review #2** - Data quality check including<br><br>1. data privacy and leakage<br>2. sampling and bias<br>3. labeling quality |
| | **CS #3**<br><br>• Complicated model developmental process and model specification, including pretrained model vs fine tuning.<br>• Prompt input by users for generative models can introduce risk if not designed appropriately | **CS Review #3** –Model specification, including prompt engineering/tuning and Hyper-parameter tuning |
| | **CS #4** - Lack of transparency and trust | **CS Review #4** - Model explainability |
| | **CS #5** - Potential bias especially when unstructured data is used. Individuals of minority group may be adversely impacted. | **CS Review #5** -Fairness testing |
| | **CS #6** Justification of the selected framework and its final specification | **CS Review #6** - Benchmarking |
| Outcomes analysis (OA) | **OA #1**<br><br>• Model output is usually unstructured data, and the complexity of output results in more measurement perspectives per different scoring factors.<br>• Randomness introduced in the decoder of generative models can prevent the output being 100% reproducible. | **OA Review #1** - Performance metrics and model output replication |
| | **OA #2** - Lack of generalization due to the complicated model structure with large size of parameters. | **OA Review #2** - Model robustness and overfitting |
| | **OA #3**<br><br>• Model may not perform as intended under certain specific or stressed circumstance under which data pattern was insufficiently observed during the model developmental process.<br>• High risk of adversarial attack | **OA Review #3** -Detect model error patterns.<br><br>Detect model performance weakness/vulnerability under stressed scenarios. |
| | **OA #4** - Generative models can generate factually incorrect output. | **OA Review #4** – Hallucination detection |
| | **OA #5** - Generative models can generate output that is complete, relevant, and factually correct but is misaligned to the decorum of acceptable language. | **OA Review #5** – Toxicity detection |

## Conceptual Soundness

### CS #1 (Literature Review)

GenAI models typically build upon Foundation Models (FM) and the risk profile of the FM must be evaluated for the target business use case. This is a widely discussed topic which not only involves discussing the abilities and weaknesses of the models (Bommasani, et al. 2021) but also their social impact (Kapoor, et al. 2024). Therefore, it is paramount to include a literature review focusing on a thorough evaluation of the proposed FM to be used in the modeling approach.

### CS #2 (Data Quality)

While certain aspects of the data quality such as scrubbing for sensitive information, annotation quality and population consistency between train and hold-out data are consistent with traditional predictive models using text, there are other unique challenges with GenAI model's data quality checks. We briefly discuss this in context of GenAI models below.

1. **Data Privacy and Security** - Data privacy and leakage require attention at both the stage of model development/fine-tuning and at implementation (Das, Amini and Wu 2024) (Cohen, Bitton and Nassi 2024). Techniques such as jailbreaking prompts can be utilized to manipulate models to bypass safety restrictions put in place by the developers and reveal unauthorized information. Hence, tests should be designed to identify the presence of any Personally Identifiable Information (PII) or any other confidential material that needs to be masked for policy or regulatory reasons.  RegEx (Regular Expression) suite  (Venkatesh 2021) can be leveraged to detect formatted PII - credit card numbers, social security numbers (SSN) and emails. Lastly, extensive testing should be performed on the model to prevent any jailbreaking [ (Peng, et al. 2024) (Microsoft Threat Intelligence 2024)].
2. **Data Sampling and Bias**: Generative AI model evaluation samples involve a textual context and a generative output that must be manually reviewed across several performance dimensions. It is paramount that these sample sizes are representative of the population (including adequate sample size devoid of any biases) in order to elicit meaningful conclusions. In absence of meta-data that allows a natural population segmentation, a common strategy involves sampling methodologies entailing embedding models (to represent unstructured data) to cluster the population for deriving a meaningful stratified sample.
3. **Annotation Quality** - In cases involving generative summaries or content generation, if the model is finetuned on ideal summaries or ideal content templates, a thorough review is necessary to certify a) the quality of these examples, b) annotation/quality consistency, and c) adequate sample representation from each annotation class.

### CS #3 (Model Specification)

All generative AI models are customized from underlying FM that are either fine-tuned on domain specific data or are used out of the box with specialized prompts. Therefore, the model specification must include a deep discussion on the choice and the parameters of customization. This includes rationale for hyper-parameter selection and loss functions for the purpose of fine-tuning and the choice of prompts for prompt tuning. Regardless, a rationale for choice of decoding parameters [ (Bengesi, et al. 2024) (HuggingFace n.d.)] must be discussed. Further, most FM include an inherent limitation around the size of the input context and that constraint must be evaluated in context of the domain specific input data.

### CS #4 (Model Explainability)

Model explainability is challenging for generative AI models where input and output are both texts. Traditional methods such as SHAP (Lundberg and Lee 2017) and LIME (Ribeiro, Singh and Guestrin 2016) may still be applied but they suffer from several challenges including computational complexity and/or restricting input features to tokens (or words). Methods confirming the relationships between the semantics of the input context and output generation are required, for example efforts to establish context source for each fact in a generation. However, this direction tends to address the issues with local explainability for each input-output pair. Global explainability is a significantly harder problem and perhaps requires a paradigm shift in how explainability is supposed to be defined for generative models. Perhaps behavioral testing including the choice of evaluation sets (Ribeiro, Wu, et al. 2020) to profile model outputs and generalize behavior could be an interesting direction. Considerable research is required in this area.

### CS #5 (Bias and Fairness)

Generative AI models are customized using either fine-tuning or out-of-box prompt tuning on FM. FM are trained on massive public datasets that contain inherent biases that may propagate into downstream customized use cases and result in unfair treatment of protected demographic groups based on sex age, race, etc. It is virtually impossible to curate the datasets for FM and control for such biases. Therefore, the only practical solution involves applying a sleuth of guardrail models on the model output to detect bias and ensure fairness. Similar guardrail models may also be applied during validation to measure the bias in the model output. Quantification of bias in model output can be performed either through observing metrics in protected vs non-protected groups or by evaluating the models on benchmark datasets for bias evaluation. (May, et al. 2019) (Caliskan, Bryson and Narayanan 2017) (Gallegos, et al. 2023).

### CS #6 (Benchmarking)

Benchmarking is a mechanism to demonstrate a choice of an alternate modeling methodology (involving an alternate architecture or FM or prompt template, etc.) that either alleviates model weaknesses in the original model object or proposes an alternate simplified framework achieving a similar objective. The second choice is of particular interest for Generative AI models. Simpler methods (say extractive summarization in place of generative summarization) must be evaluated to justify the additional complexity for a generative model. Of course, it is totally acceptable to use higher parameter models with or without fine-tuning as a benchmark to demonstrate efficacy. It must however be noted that choice of a computationally intensive model as a means of remediating model weaknesses may not always be feasible given the costs and complexity of deploying such models.

## Outcome Analysis

### OA #1 (Performance evaluation and replication)

The auto-regressive nature of the current GenAI models pose additional challenges in replicating model outputs. In addition to reproducing computational environments, all decoding parameters including parameters for random number generation algorithms must be documented. In absence of exact reproducibility, uncertainty in generated output should be quantified using techniques such as semantic invariance. (Kuhn, Gal and Farquhar 2023) (Lin, Trivedi and Sun 2023)

Next, the unstructured GenAI model output often requires evaluation across several performance dimensions. These evaluation dimensions (metrics) are often task dependent. For example,

<u>Summarization</u>: Evaluations must be made along the dimensions of

- Completeness - generation captured all key points.
- Hallucinations - generation did not misrepresent facts (discussed in detail under OA#3)
- Fluency - generation was coherent and grammatically correct.

<u>Retrieval Augmented Generation</u>: (Es, et al. 2023). Evaluations must be made along dimensions of

- Faithfulness – how many facts (statements) in the generation are supported by retrieved context.
- Answer relevance – how closely the generation aligns with query in terms of completeness and redundancy.
- Context Relevance – how closely the retrieved context aligns with the query in terms of sufficiency and exclusivity of information.

Evaluation across several such dimensions where inputs and outputs are both texts can turn out to be prohibitively expensive if solely relied on human annotations as is done in case of predictive models. This requires development of scalable automated metrics which is an active area of research encompassing traditional NLP techniques such as NLI (Bowman, et al. 2015) (Williams, Nangia and Bowman 2017)to using LLMs as a judge [ (Es, et al. 2023)]. However, these automated metrics would suffer from their own inadequacies and would require calibration to human judgments to ascertain their error bounds (Sudjianto, et al. 2024) and simultaneously draw performance conclusions at a meaningful sample size.

## OA #2 (Robustness Testing for Model Generalization)

Any machine learning model including GenAI models require evidence of generalization (Barbiero, Squillero and Tonda 2020) (H. Li, et al. 2020). While the evidence on the holdout dataset serves as one measure, typically, for text models, robustness testing is employed to gain additional confidence on model's generalization. Robustness testing involves controlled perturbations to the input text with an expected output. The degree of deviation of the model's output from the expected output in response to perturbations is generally aggregated to define generalization measures. These simple ideas can readily be applied to the key generative model uses such as summarization or RAG. For example, in summarization, the input context can be perturbed such that it retains its semantics (through synonym replacement or introduction of misspellings), and the output may be checked for semantic similarity with the original output. For RAG, the perturbations may be applied at various stages – the input query or within the retrieved context (as simple as changing the order of the retrieval or injecting semantic preserving changes).

## OA #3 Weakness Detection

Weakness detection entails finding segments within the input population where model performance is unacceptable. For text data, such segments may be purely defined in terms of their semantics or their linguistic characteristics. Embedding models (that capture the targeted segmentation profile) are often used in conjunction with clustering algorithms (Grootendorst 2022) such that model performance is measured over different meaningful clusters to identify weak regions (Li, Singh, et al. 2024).

**OA #4 Hallucination**

Hallucination: LLMs may generate outputs that appear plausible but are incorrect or nonsensical. This phenomenon is referred to as hallucination and becomes a challenge when relying on LLMs for generating accurate and reliable content. Thus, detecting hallucination for a specific LLM application is critical to assure reliability and trustworthiness of its generated output.

From a model validation perspective, several automated methods may be used to detect the presence of hallucinations in the output. We list a few of these approaches below:

1) Natural Language Inference (Bowman, et al. 2015): Utilizes a hypothesis and inference setup to evaluate factual consistency between generated content against the context.
    - Set the source context as a premise and a chunk of the generated output as hypothesis.
    - Convert the logit value of classifying the hypothesis as a contradiction to a hallucination score.
2) Self-check GPT (Manakul, Liusie and Gales 2023): this technique is based on the uncertainty produced by LLM models and utilizes the stochastic inference capabilities to generate multiple outputs. The idea is utilizing inconsistency in multiple generations as indicator for hallucination.
    - Use LLM to re-generate multiple outputs, with the same query/prompt.
    - Compare consistency between (a) original output and (b) re-generated outputs.
3) The second approach is based on fact check, and inspired by Chain-of-Verification (Dhuliawala, et al. 2023):
    - Use LLM to identify key facts in output, then generate verification questions accordingly.
    - Use LLM to generate answers to verification questions independently.
    - Measure consistency between (a) original output and (b) answers to verification questions.

For each approach, a score indicating the degree of hallucination is calculated for each generated output which are then summarized to assess a LLM application's overall risk for hallucination. However, such LLM-based hallucination metric, while allowing scalable calculation, may contain inaccuracies. Therefore, it is always necessary to draw a sample and conduct human evaluation, to assure quality and establish error bound on the LLM-based metric.

**OA #5 Toxicity**

Toxicity is the general alignment [ (Askell, et al. 2021)] problem in generative AI models where factually correct and complete outputs may contain language misaligned with the decorum set for communications. At model development stage, toxicity may be controlled through instruction fine-tuning (Tay, et al. 2022) (Hawkins, Mittelstadt and Russell 2024), Reinforcement Learning with Human Feedback (RLHF) (Wu, et al. 2023) and even prompt tuning (Welbl, et al. 2021). However, at model implementation stage, specialized models trained to detect toxic content (guardrails) need to be used to prevent toxic content being exposed to users or customers (Hanu, Thewlis and Haco 2021). Therefore, at model validation stage as well, a large set of model outputs need to be tested using similar such models (Davidson, et al. 2017) and restrictions may be imposed on the use GenAI models with unacceptable degree of toxicity without appropriate guardrail models in production.

# Implementation Testing and Model Use

Implementation of all models require standard model-level control activities such as controls on model data input and integrity, model output control (avoid errors in output), controls over model change management (such as identification and logging of change events, version control process, review and approval process, compensating controls to address model weakness or limitations, etc.) and access management. At time of model development, developers provide all evidence of implementation testing including model functions accessed by implementation process, data sources used as input, model configurations, description of production platform, known gaps in ongoing model maintenance and date and version of implementation along with evidence of all required controls in place.

GenAI models are associated with heightened risk discussed in Table 2-2 and hence require additional scrutiny in the form of added model level controls depending on the model use and model user. Note that any model use suffers from heightened reputational risk if they are directly used, or its output is directly exposed to customers. Hence, the financial institution may prescribe increased or enhanced controls based on the risk determined from the model usage and user. Some of the common forms of GenAI control are summarized in Table 3-2.

Model implementation testing needs to be carried out not only before start of model use but also intermittently to ensure that the model is functioning as intended. Additionally, implementation should be tested in case of any change to model, its data source, or the model implementation environment.

At the stage of model validation, all evidence of implementation, testing, and control should be reviewed. Development environment and production (model use) environment may be different and produce inconsistencies in model outcome given the same prompt and same decoding parameter. Hence validators may independently test for output consistency in the two environments.

**Table 3-2: Forms of Control with examples**

| Controls | Description | Examples |
|---|---|---|
| **User Control** | Ensure only authorized users use the model appropriately. | - Limiting user access through access management IT controls.<br>- Users receive certified training for heightened risks prior to using the model |
| **Usage Control** | Ensure the model is used responsibly only for its intended purpose. | - In implementation, lock down the model functionality to only allow for a specific approved usage(task).<br>- Maintain records of all or a sample of model interactions to monitor compliance and investigate misuse |
| **Human-in-the-loop control** | Ensure that the model output is not fed into any automated workflow | - Model outputs are always reviewed by certified humans with subject matter expertise prior to being used for any decisions. |

| 'Terms of use' alert control | Ensure the model user is aware of all heightened risks associated with the model use. | <ul><li>Embed "Term of use" alerts in user interface for GenAI platforms.</li><li>Include disclaimers within the generated text or alongside it.</li></ul> |
|---|---|---|
| Input Control | Ensure the model is not queried with queries where outputs would manifest the heightened risks. | <ul><li>Use preprocessing to block harmful or unethical input prompts.</li><li>Define a collection of pre-designed prompts, templates or examples curated to guide the behavior of GenAI</li></ul> |
| Output Control | Ensure the output is screened for heightened risks prior to use. | <ul><li>Guardrails (typically additional models) to detect and minimize toxicity and hallucination risks.</li><li>Set bounds on response length or complexity to avoid generating overly detailed or harmful outputs.</li></ul> |

The mitigation of risks associated with GenAI models is not limited to the implementation of above forms of control and must encompass additional procedural checks in all stages of model risk management starting with initial risk assessment to implementation testing and finally model output monitoring.

## Ongoing monitoring

The model risk related to any model use needs to be monitored periodically based on an established monitoring plan that includes:

- A reasonable monitoring frequency.
- A justified monitoring approach.
- Sufficient and clearly defined Key Performing Indicator (KPI) metrics that address key model risks.
- Acceptable thresholds for the KPIs that can successfully capture any deteriorating model performance.
- A final decision-making guidance based on the multiple KPIs.
- Action plans for poor model performance.

A well-developed monitoring plan should reveal error patterns that lead to additional insights on model weakness.

GenAI model monitoring plans may require additional model KPIs to monitor toxicity and hallucination risks, query domain stability, etc. or operational KPIs relevant to LLM/GenAI model testing such as real time user-feedback, number of successful vs attempted generation, etc. Moreover, automated metrics in conjunction with human calibration might be required to alleviate costs of human annotations.

Model validation activities can be leveraged to review the monitoring plan for adequacy and review latest model monitoring results to determine whether model is performing as expected.

# 4 Conclusion

This paper presents a model risk framework for managing model risk within GenAI models with a special emphasis on the additional testing required at the model validation stage. We realize that these models and their use cases are in an early stage, and we expect these use cases to evolve over the next few years. Regulations for GenAI are also expected to evolve and would require updating the model risk framework. It promises to be an exciting journey, and the authors expect this framework to undergo substantial changes in the future.

# Appendix A

## Terminology

The terminologies used within this paper are defined as follows:

**Artificial Intelligence (AI)** refers to systems used to automate tasks that historically required human intelligence. These are usually characterized by complex inputs, unstructured data and/or sophisticated decision making.

An **LLM** is a language model (LM) notable for its ability to achieve general-purpose language understanding and generation and is characterized by the large size of its parameters. LLMs are artificial neural networks which are pretrained using self-supervised learning and semi-supervised learning[1]. LLMs largely represent a class of deep learning architectures called transformer networks[2]. Typically, pre-trained LLMs are also commonly referred to as **foundation models (FM)** implying that they can be used as a start point and customized into a domain specific model.

**GenAI** is a type of AI system capable of generating new content including text, images, and videos. This paper focusses on GenAI producing text output which use LLM based FM as their building block.

---

[1] https://en.wikipedia.org/wiki/Large_language_model
[2] Transformers were first introduced by Google in the 2017 paper "Attention Is All You Need" (https://arxiv.org/abs/1706.03762)